# Concurrent Auctions Across The Supply Chain

**Moshe Babaioff** MOSHEB@CS.HUJI.AC.IL
**Noam Nisan** NOAM@CS.HUJI.AC.IL
*School of Computer Science and Engineering,*
*The Hebrew University of Jerusalem, Jerusalem 91904, Israel*

## Abstract

With the recent technological feasibility of electronic commerce over the Internet, much attention has been given to the design of electronic markets for various types of electronically-tradable goods. Such markets, however, will normally need to function in some relationship with markets for other related goods, usually those downstream or upstream in the supply chain. Thus, for example, an electronic market for rubber tires for trucks will likely need to be strongly influenced by the rubber market as well as by the truck market.

In this paper we design protocols for exchange of information between a sequence of markets along a single supply chain. These protocols allow each of these markets to function separately, while the information exchanged ensures efficient global behavior across the supply chain. Each market that forms a link in the supply chain operates as a double auction, where the bids on one side of the double auction come from bidders in the corresponding segment of the industry, and the bids on the other side are synthetically generated by the protocol to express the combined information from all other links in the chain. The double auctions in each of the markets can be of several types, and we study several variants of incentive compatible double auctions, comparing them in terms of their efficiency and of the market revenue.

## 1. Introduction

The recent rush towards electronic commerce over the Internet raises many challenges, both technological and conceptual. This paper deals with the conceptual challenge of coordination between electronic markets. Let us look only a few years into the technological future of electronic commerce. It seems very likely that the following two key challenges will be adequately solved by the industry:

- **Supply Chain Integration:** The enterprise information systems of businesses will be able to securely and efficiently share information and inter operate with the information systems of their suppliers, customers, and partners.
- **Electronic Markets:** Efficient, sophisticated, robust and liquid electronic markets will be available for the trade of goods in most segments of the industry. Such markets will interactively respond to changes in supply and demand, dynamically changing trade quantities and prices.

We are interested in the conceptual question of how can markets for related goods share information. Consider, for example, a fictional market for rubber tires for trucks, and the two related markets for rubber and for trucks. One can imagine the following simplified supply chain forming: rubber manufacturers placing sell bids for rubber in the rubber market; tire manufacturers placing buy orders on the rubber market and sell bids on the tire market; truck manufacturers placing buy bids in the tire market and selling trucks on the truck market; and finally customers bidding for trucks. One would expect the combination of these markets together with the information systems of the manufacturers to be able to automatically respond to markets changes in an economically efficient way. Thus, for example, a surge in demand for a certain type of trucks, will raise their price in the truck market causing manufacturers of this type of truck to automatically decide to increase production, consequently and automatically raising their electronic bids for tires. This, in turn, may increase tire prices in the tire market, etc., etc., leading, eventually and indirectly, but still completely automatically, to increased rubber production by rubber manufacturers.

Let us emphasize: the process just described occurs, slowly, in normal human trade by the combined effects of a large number of self-interested decisions by the many people involved in this supply chain. What we desire from the combination of the participating information systems and electronic markets is to automatically (without human control) and very rapidly (within the time-frames of electronic commerce), to reach similar results or even more economically efficient ones than what humans usually achieve. These results should be achieved despite the fact that the information systems of manufacturers will still be self-interested – optimizing the company's profit and not the global economic efficiency.

Seeing the "invisible hand" function in normal human economic activity, one would certainly expect that electronic markets reach these types of results. However, a key conceptual design challenge emerges when bidders must be *concurrently active in more than one market*. Tire manufacturers must concurrently participate as buyers in the rubber market and as sellers in the tire market. Clearly, the quantity of rubber they wish to buy at a given price is determined by the amount of tires they can sell at a given price. Thus, the price they bid for buying rubber must be intimately related to the price they bid for selling tires – any increase in one of them will lead to a corresponding increase in the other. It is theoretically impossible to define a suggested bid for one without the other. Thus, if the two markets operate independently, then the tires manufacturers are not able to reasonably participate in any of them. If they operate sequentially, say first rubber is bought and only afterwards tires are sold, then a serious exposure problem emerges: tire manufacturers must be conservative in their bids for rubber, as they do not know in advance what price they will get in the tire market.

One approach for handling this inter-dependence of markets is to run the complete supply chain as a single complex huge market. Conceptually this is in the spirit of the recently popular "vertical markets" and "vertical portals" that try to vertically integrate information and trade for a complete vertical segment of industry. The integration of all these markets into a single complex market results in a complex optimization problem, and some research has been done to address such problems as pure optimization problems. Such a centralized solution has obvious advantages, but is also problematic due to necessity of concentrating all information, communications, and decision making at a single point. Such centralization of information is problematic both in the sense of distributed computing systems and in the economic sense.

In this paper we suggest an alternative approach: the supply chain is organized as a sequence of separate markets (which do not act strategically) that communicate among themselves using a fixed protocol to create a distributed mechanism (for a general overview of distributed algorithmic mechanism design, we refer the reader to Feigenbaum & Shenker, 2002). A similar approach has been suggested by Walsh and Wellman (2003) and by Walsh, Wellman, and Ygge (2000), who formulate a general problem that is NP-complete and obtain solutions that are not provably efficient, either in the computational sense or in the economic sense. A much simpler problem of a linear supply chain was considered by Bosch-Domenech and Sunder (1999), but no provably efficient protocol was suggested. We also consider the linear supply chain problem, and obtain computationally efficient protocols which achieve provably high economic efficiency with budget balance (or full efficiency with budget deficit). In our protocols, the intermediate markets along the chain transform one good into another. Thus, for example, tire manufacturers place bids for the operation of "transforming a unit of rubber into a tire". The protocol between the different markets assures that the different markets reach compatible decisions, i.e. that the amount of "rubber units to tires" that was allocated is equal both to the amount of rubber manufactured and to the amount of tires needed. Furthermore, these amounts achieve global economic efficiency across the supply chain. Finally, this result does not assume that the manufactures information systems have any global knowledge or behavior, beyond their knowledge of their own cost structure and their self-interested (rational) behavior.

This paper focuses on the case of a simple linear supply chain for discrete units of goods, where each manufacturer is able to transform a single unit of one good into a single unit of another good, incurring some cost for the transformation. Each consumer obtains value from acquiring a single unit of the final good. We assume that each agent has a quasi-linear utility function, and his goal is to maximize his utility.

Each of our markets takes the form of a double auction (see Friedman & Rust, 1991, for a study of double auction markets), and we consider several variants of double auctions. These variants address four issues: *Incentive Compatibility (IC):* The double auction rules motivate self-interested agents (the manufacturers and consumers) to reveal their costs / values truthfully (we use the standard notions of dominant strategies IC from Mechanism Design literature; Mas-Collel, Whinston, & Green, 1995; Osborne & Rubinstein, 1994). *Individual Rationality (IR):* Every agent has a strategy which ensures a non-negative utility, so agents participate voluntarily (Ex-Post individual rationality). *Economic Efficiency:* The desired outcome should optimize the sum of valuations of all participants. *Budget Balance (BB):* The payment by each buyer does not necessarily equal to the payment received by each seller, but we wish to ensure that the market mechanism itself does not subsidize the trade. Thus, the total payment by the buyers should be at least as the total amount given to sellers. By Myerson and Satterthwaite (1983) impossibility result, these four conditions cannot apply simultaneously. The paper by Parkes, Kalagnanam, and Eso (2001) presents mechanisms for combinatorial exchange which are BB and IR but fairly efficient and fairly incentive compatible. We take a different approach and consider variants that are always IR and IC and trade-off the last two conditions of efficiency and budget balance. Such a deterministic rule was previously suggested by McAfee (1992), but, surprisingly, this rule turned out not to be compatible with our supply-chain protocols. Therefore we suggest two new randomized double auction rules that can be used with our supply-chain protocols, and obtain budget balance or surplus but with a slight loss of efficiency. We also provide some simulation results comparing the efficiency and budget surplus of the different variants of double auctions.

The main contribution of this paper is the description of two alternative protocols that allow a supply chain mechanism formed from double auctions to operate efficiently. These protocols are computationally efficient in terms of communication and computation time. We prove that when these protocols are applied with a specific double auction rule (the Trade Reduction rule), the resulting system exhibits the following three key properties:

1. Ex-post Individual Rationality and Incentive Compatibility (in the sense of dominant strategies).
2. High global economic efficiency (not optimal, but as good as the underlying double auction which guarantees high fraction of the efficient allocation value).
3. Budget Balance.

With our new double auctions, an even higher global economic efficiency can be achieved in expectation, but with expected (ex-ante) budget balance rather than worst case (ex-post). When these protocols use the VCG (Vickrey, 1961; Clarke, 1971; Groves, 1973) double auction rule which is efficient, they maintain IR and IC and achieve full economic efficiency for the entire supply chain, but with a budget deficit.

Clearly our stylized model is not the most general supply chain model and it is far too simple to be a realistic model. Our purpose in analyzing this model is to present a first step towards a mechanism design approach to supply chain problems. As far as we know we are the first to design a mechanism that is IC, IR, BB and highly efficient for a supply chain model.

The rest of this paper is structured as follows. In Section 2 we give a complete self contained example of the organization of a simple supply chain and we demonstrate the type of calculations and information transfer of one of our protocols. In Section 3 we present the model and conditions for incentive compatibility. In Section 4 we summarize the properties of several variants of incentive-compatible double auctions and present two new randomized double auction rules. In Section 5 we present two alternative protocols for supply chain coordination between markets, and prove the properties achieved by these protocols, and in Section 6 we conclude with directions for future work.

## 2. The Lemonade-Stand Industry

Charlie Brown has decided to draw on his vast experience in the lemonade stand industry and transform the whole industry by bringing it online to the Internet. Charlie Brown has already struck partnerships with strategic players from the three core segments of the industry:

- **Lemon Pickers:** Alice, Ann, and Abe can pick a lemon from the neighborhood lemon tree (one lemon maximum per day).
- **Lemonade Squeezers:** Bob, Barb, and Boris know how to squeeze a single lemon and make a glass of lemonade from it (one glass maximum per day).

- **Lemonade Consumers:** Chris, Carol, and Cindy want to buy one glass of lemonade each (per day).

Charlie Brown has obtained a preliminary version of this paper and has built his Internet systems accordingly, using the *Symmetric Protocol* suggested here. Charlie Brown has created three communicating electronic markets: A *lemon market* through which Alice, Anna, and Abe will sell lemons. A *squeezing market* in which Bob, Barb, and Boris will offer their squeezing services, and a *juice market* in which Chris , Carol, and Cindy can buy lemonade.

In the first day of operations each of the participants logged into his or her market and entered a bid: Alice was asking for $3 in order to pick a lemon, while Ann wanted $6, and Abe (who was living farthest from the tree) wanted $7. Bob, Barb, and Boris were asking for, respectively, $1, $3, and $6 in order to squeeze a lemon, while Chris, Carol, and Cindy, were willing to pay, respectively, $12, $11, and $7 for their glass of lemonade. Figure 1 presents the three markets, the supply curves in the lemon market ($S^L$) and the squeezing market ($S^{L\to J}$), and the demand curve in the juice market ($D^J$). Knowing that the auction works such that his optimal bidding strategy is to report his true value or cost, each agent reports truthfully. Let us follow the operation of the system and see how it manages to reach socially efficient allocation and decide how many lemons should be picked to be squeezed into lemonade.

| **Lemon Market** | |
|---|---|
| **Supply** ($S^L$) | **Demand** ($D^L$) |
| **3** | |
| **6** | |
| **7** | |

| **Squeezing Market** | |
|---|---|
| **Supply** ($S^{L\to J}$) | **Demand** ($D^{L\to J}$) |
| **1** | |
| **3** | |
| **6** | |

| **Juice Market** | |
|---|---|
| **Supply** ($S^J$) | **Demand** ($D^J$) |
| | **12** |
| | **11** |
| | **7** |

Figure 1: The Supply Chain bids

In the first stage, the markets send information to each other in two phases. In the first phase, the lemon market aggregates the supply curve for lemons, $S^L$, and sends this information to the squeezing market. The squeezing market aggregates the supply curve for squeezing services, $S^{L\to J}$, adds this vector, point-wise, to $S^L$, and sends the sum to the juice market. For the juice market, this sum represents the supply curve for juice, $S^J$, aggregated over the complete supply chain. As can be seen in Figure 2, the cost of the first glass of lemonade is $4, since the first lemon cost is $3 and the squeezing operation cost is $1.

In the second phase, the juice market sends the demand curve for juice, $D^J$ to the squeezing market, which subtracts from it, point wise, the supply curve for squeezing services, sending the difference vector to the lemon market, where this is interpreted as the demand curve for lemons, $D^L$, aggregated over the complete supply chain. As can be seen in Figure 3, the demand for the first lemon is $11, since the first glass of lemonade has a demand of $12 but the squeezing operation cost is $1.

The net demand curve for squeezing services, $D^{L\to J}$ can now be calculated by the squeezing market to be $D^J - S^L$. As can be seen in Figure 4, the demand for the first

| Lemon Market | | Squeezing Market | | Juice Market | |
|---|---|---|---|---|---|
| **Supply** ($S^L$) | **Demand** ($D^L$) | **Supply** ($S^{L\to J}$) | **Demand** ($D^{L\to J}$) | **Supply** ($S^J$) | **Demand** ($D^J$) |
| **3** | | **1** | | 4 | **12** |
| **6** | | **3** | | 9 | **11** |
| **7** | | **6** | | 13 | **7** |

$\xrightarrow{S^L}$ $\xrightarrow{S^L+S^{L\to J}}$

Figure 2: The Supply Chain after supply graphs propagation

| Lemon Market | | Squeezing Market | | Juice Market | |
|---|---|---|---|---|---|
| **Supply** ($S^L$) | **Demand** ($D^L$) | **Supply** ($S^{L\to J}$) | **Demand** ($D^{L\to J}$) | **Supply** ($S^J$) | **Demand** ($D^J$) |
| **3** | 11 | **1** | | 4 | **12** |
| **6** | 8 | **3** | | 9 | **11** |
| **7** | 1 | **6** | | 13 | **7** |

$\xleftarrow{D^J-S^{L\to J}}$ $\xleftarrow{D^J}$

Figure 3: The Supply Chain after demand graphs propogation

squeezing operation is $9 since the first glass of lemonade has a demand of $12 but the first lemon has a cost of $3. This means that if the lowest cost of squeezing operation is less than $9, at least one glass of lemonade can be manufactured.

| Lemon Market | | Squeezing Market | | Juice Market | |
|---|---|---|---|---|---|
| **Supply** ($S^L$) | **Demand** ($D^L$) | **Supply** ($S^{L\to J}$) | **Demand** ($D^{L\to J}$) | **Supply** ($S^J$) | **Demand** ($D^J$) |
| **3** | 11 | **1** | 9 | 4 | **12** |
| **6** | 8 | **3** | 5 | 9 | **11** |
| **7** | 1 | **6** | 0 | 13 | **7** |

Figure 4: The Supply Chain after constructing the supply and demand graphs

At this point all three markets have both a supply curve and a demand curve, and each market can conduct a double auction.

Being an Internet startup, Charlie Brown has decided to subsidize the trade in his markets, ignoring the sections of this paper that aim to eliminate any budget deficit of the markets. In each market he thus uses the VCG (Vickrey, 1961; Clarke, 1971; Groves, 1973) double auction rule[1], derived from the Vickrey, Clarke and Groves general auction scheme. The VCG double auction picks the highest value allocation in each market, so the winners are the agents in the socially efficient allocation. The efficient allocation in each market includes the two highest value (lowest cost) bidders, since each of the two trades has a positive gain, and the third trade has a negative gain. The VCG double auction charges

1. A formal definition of the VCG double auction rule appears in Section 4

each consumer (demand side) the minimal value the agent must bid in order to be in the efficient allocation. Similarly, each supplier (supply side) receives a payment that equals the maximal cost it may bid and still be in the efficient allocation. The payment takes into account both the competition with other agents of the same class and competitiveness of the bid with respect to bids of agents of the other classes. With that payment scheme, the best strategy of each agent is to bid truthfully. In the lemon market, two lemons are sold (by Alice and Ann) for $\$7 = min(\$7, \$8)$ each (in order to win, each must bid a cost lower than the cost of Abe, the first non-winning lemon supplier, which is $7. Each must also bid a cost lower than $8, otherwise the cost of her lemon will be too high to match the lowest demand for a lemon). In the squeezing market, two squeezing contracts are awarded (to Bob and Barb) for $\$5 = min(\$6, \$5)$ each (since if one bids a cost higher than $5, there will be no demand for his squeezing operation, and the squeezing operation cost of Boris is $6, which is higher than $5). In the juice market, the VCG rule awards two glasses of lemonade (to Chris and Carol) for the price of $\$9 = max(\$7, \$9)$ each (since they must bid at least $9 to match the supply of juice, and if they do, they also defeat Cindy).

Charlie Brown is thrilled: the different markets have all reached the same allocation amount, 2, which, he has verified is indeed the social optimum: Society's net gains from trade in his system are $(12 + 11) - (1 + 3) - (3 + 6) = \$10$, which can't be beaten. Charlie Browns' investors are somewhat worried by the fact that the system subsidized every glass of lemonade by \$3 ($= 7 + 5 - 9$), but Charlie Brown assures them that changing the double-auction rules to one of the other double auction rules suggested in this paper can lead to a budget balance or even surplus, while maintaining high social gain. The nine trading partners have evaluated carefully the operation of this chain of markets and have assured themselves that they are best served by always bidding their true cost structure.

## 3. IC Mechanisms for Single-Minded Agents

Section 3.1 presents an abstract model for agents that partition the set of outcomes to two, the case they win and the case they loss. Each agent has some valuation for the case he wins, and this valuation is represented by a single parameter. The double auction model and the supply chain model we consider are special cases of this model. The main theorem in this section presents necessary and sufficient conditions for incentive compatibility in this abstract model. Mu'alem and Nisan (2002) have presented this theorem in the context of combinatorial auction with agents that desire one specific bundle that is publicly known (the Known Single Minded Model). The theorem is derived from a more general result for agents with private bundles (the Single Minded Model), proved by Lehmann, O'Callaghan, and Shoham (2002). Similar result for a different general model with one-parameter agents by Archer and Tardos (2001) appears in the Computer Science literature. For completeness, we present the results that are relevant to our single-minded agents model.

In section 3.2 we farther restrict the model of the single-minded agents to the case of sets of agents that can replace each other, and characterize non-discriminating mechanisms for that case.

### 3.1 General Model

In the *Single-Minded agents model*, there is a finite set of agents $N$ and a set of outcomes $O$. For each agent $i \in N$, $O$ is partitioned to two disjoint sets $O_i^W$ and $O_i^L$, the outcomes in which he Wins and the outcomes in which he Loses, respectively (this partition is public knowledge). For any $i$ there is an ordered set of *values space* $V_i$ for the case he wins. Agent $i$ has a private *value* $v_i \in V_i$ for any $o \in O_i^W$ (he wins) and a value of 0 for $o \in O_i^L$ (he loses). The private values are the only private information, all other information is publicly known to all participants and to the mechanism (and it is known that the rest of the information is public). For technical reasons we assume that for any $v_i^1, v_i^2 \in V_i$, such that $v_i^1 > v_i^2$, there exist $v_i^3 \in V_i$ such that $v_i^1 > v_i^3 > v_i^2$ (For example $V_i$ can be the set of Real numbers or Rational numbers). When agent $i$ wins and pays $p_i$ he has a quasi-linear *utility* $u_i = v_i - p_i$, and in normalized mechanisms he pays 0 and has 0 utility if he loses. We assume that agents are self-interested and try to maximize their utility. Under this model, an outcome $o \in O$ has a one-to-one mapping to a set of winners and is called an *allocation* $A$, $A = \{i \in N | o \in O_i^W\}$. We say that the allocation $A$ is *efficient* if $\sum_{i \in A} v_i$ is maximized. The *efficiency* of the allocation $\hat{A}$ is $\frac{\sum_{i \in \hat{A}} v_i}{\sum_{i \in A} v_i}$, where $A$ is an efficient allocation. Let $V$ be the set of possible values of the agents, $V = \prod_{i \in N} V_i$. We work with mechanisms in which agents are required to report their values, and the mechanism decides the allocation and the payments to the agents in a deterministic way. The reported value $b_i \in V_i$ of agent $i$ is called the *bid* of the agent and might be different from his private value $v_i$. Let $b \in V$ be the bids of all agents. An *allocation rule* $R$ decides the allocation according to the reported values $b \in V$, $R$ is a function $R : V \to O$. A *payment rule* $P$ decides the payment $p_i$ of agent $i$, $P$ is a function $P : V \to R^N$. A *mechanism* $M$ defines an allocation and a payment rule, $M = (R, P)$. Mechanism $M$ is *Budget Balanced (BB)* if $\sum_i p_i \geq 0$ for any bids $b \in V$. $M$ is *Incentive-Compatible (IC) in dominant strategies* if for any agent $i$, bidding $v_i$ maximizes $i$'s utility over all possible bids of the other agents. $M$ is *normalized* if losing agents have a payment (and utility) of 0. $M$ is (ex-post) *Individually Rational (IR)* if for any agent $i$ and value $v_i$, there is a bid $b_i$ such that when he bids $b_i$, $u_i \geq 0$ for all possible bids of the other agents. Note that any normalized and incentive-compatible mechanism is individually rational (since truthful bidding ensures a non negative utility).

Below we present necessary and sufficient conditions for a mechanism to be incentive compatible in dominant strategies under the single-minded agents model. These conditions will later be used to prove properties of our double auctions and supply chain mechanisms created by our protocols.

For bids $b \in V$ we denote $b = (b_i, b_{-i})$ where $b_{-i}$ is the bids of all agents but $i$.

**Definition 1.** *Allocation rule $R$ is* **Bid Monotonic** *if for any bids $b \in V$, any agent $i$ and any two possible bids of $i$, $\hat{b}_i > b_i$, if $i$ is in the allocation $R(b_i, b_{-i})$ then he is also in the allocation $R(\hat{b}_i, b_{-i})$.*

A bid monotonic allocation rule ensures that no winning agent becomes a loser by improving his bid. The following observation is a direct result from the above definition.

**Observation 3.1.** *Let $R$ be a bid monotonic allocation rule, let $b \in V$ be a set of bids of the agents and let $i$ be any agent with a bid $b_i$.*

*If there exists a bid $\hat{b_i}$ for $i$ such that $i$ is in the allocation $R(\hat{b_i}, b_{-i})$, then there exists a critical value $C_i$ such that $i$ wins if he bids $b_i > C_i$ and $i$ loses if he bids $b_i < C_i$ ($C_i$ is independent of the bid $b_i$)*[2].

Note that if the bid is equal to the critical value, the observation does not say if the agent wins or not.

**Lemma 3.2.** *Let $M$ be a normalized and IC mechanism with allocation rule $R$. Then $R$ is Bid Monotonic.*

*Proof.* Assume in contradiction that the auction is normalized and IC (thus also IR) but the allocation is not bid monotonic. Then there exists an agent $i$ and two values $b_H > b_L$, such that $i$ wins the auction and pays $P_L$ if he bids $b_L$, and loses the auction and pays zero if he bids $b_H$. The mechanism is IR therefore $b_L - P_L \geq 0$. If $i$'s true value is $b_H$ he can gain by misreporting his private value: if he reports his true value he loses the auction and has utility zero, but if he bids $b_L$ he wins the auction and pays $P_L$. In this case his utility is $b_H - P_L > b_L - P_L \geq 0$ in contradiction to the assumption that the auction is IC. □

**Theorem 3.3.** *A normalized mechanism $M$ with allocation rule $R$ is IC if and only if $R$ is Bid Monotonic and each trading agent $i$ pays his critical value $C_i$ ($p_i = C_i$).*

*Proof. Case if:* Assume that $M$ is normalized, $R$ is bid monotonic and agent $i$ with value $v_i$ pays $C_i$ if he wins. We first show that the auction is IR, we show that $i$ receives non-negative utility from bidding truthfully. If $i$ loses, he pays zero and has zero utility. If $i$ wins the auction by bidding truthfully, then by Observation 3.1, $v_i \geq C_i = p_i$, hence his utility is $v_i - p_i = v_i - C_i \geq 0$.

To prove IC we prove that $i$ cannot improve his utility by misreporting his value. Consider the case in which agent $i$ wins the auction by bidding his true value $v_i$. If $i$ bids untruthfully and loses, then he gets zero utility, which by IR cannot be better than his utility with a truthful bid. If $i$ bids untruthfully and wins the auction, then since he still pays his critical value $C_i$ his utility remains the same.

Now consider the case in which $i$ loses the auction by bidding truthfully. His utility is zero and $v_i \leq C_i$ by Observation 3.1. If $i$ bids untruthfully and loses, his utility remains zero. If $i$ bids untruthfully and wins, his utility is $v_i - C_i \leq 0$. In both cases, we have shown that an agent cannot improve his utility by bidding untruthfully, thus proving that $M$ is IC.

*Case only if:* Assume that the auction is normalized and IC (thus also IR). By Lemma 3.2 its allocation rule is bid monotonic, so we need to prove that each agent $i$ with value $v_i$ must pay his critical value $C_i$. Assume that $p_i \neq C_i$ for some agent $i$. If $p_i > C_i$, then if $p_i > v_i > C_i$ and $i$ bids truthfully, then by Observation 3.1 $i$ wins but his utility is $v_i - p_i < 0$ which contradicts individual rationality. If $p_i < C_i$, then if $p_i < v_i < C_i$ by Observation 3.1 $i$ loses and has zero utility if he bids truthfully. If $i$ misreports his value by bidding $b_i > C_i$, he would win the auction and has utility $v_i - p_i > 0$ which contradicts incentive compatibility. □

So for normalized and IC mechanisms, the allocation rule which is bid monotonic uniquely defines the critical values for all the agents and thus the payments.

2. $C_i = -\infty$ if $i$ always wins

**Observation 3.4.** *Let $M1$ and $M2$ be two normalized and IC mechanisms with the same allocation rule. Then $M1$ and $M2$ must have the same payment rule, which means that $M1$ and $M2$ are the same mechanisms.*

### 3.2 Non-Discriminating Mechanisms

We are interested in the subclass of mechanisms for single-minded agents which do not discriminate between agents that have the same "roll" in the outcome. These are agents that can always replace each other.

**Definition 2.** *Agent $i$ has the same* class *as agent $j$, if for every allocation $A$ such that $i \in A$ and $j \notin A$, the allocation $A' = A \setminus \{i\} \cup \{j\}$ is also in $O$, and for every allocation $A$ such that $i \notin A$ and $j \in A$, the allocation $A' = A \setminus \{j\} \cup \{i\}$ is also in $O$.*

Note that the class of an agent is an equivalence class.

**Definition 3.** *Let $M$ be a mechanism, $T$ be any class of agents and $i, j$ be two agents of class $T$. $M$ is* non-discriminating by identity *if for any bids $b \in V$, if $i$ wins when he bids $b_i$ and $b_j > b_i$ then $j$ also wins.*

*$M$ has* non-discriminating pricing *if for any bids $b \in V$ and every class $T$ there is a value $p_T$, such that if agent $i$ of class $T$ has a bid $b_i$ and $b_i > p_T$ then $i$ wins and pays $p_T$.*

*$M$ is* non-discriminating (ND) *if it is non-discriminating by identity and has non-discriminating pricing.*

Non-discriminating mechanisms are also called Envy-Free mechanisms by Goldberg and Hartline (2003), since no loser envies any winner for the fact that he won the auction or for the price he paid. In this paper we only consider mechanisms that are non-discriminating by identity, which means that agents of the same class are picked by their bid order, from high to low. Also note that if $M$ is IR and has non-discriminating pricing then any agent of class $T$ that bids below $p_T$ must lose the auction (otherwise he has a negative utility contradicting IR).

The following observation is a direct result of Theorem 3.3.

**Observation 3.5.** *Let normalized mechanism $M$ be IC and ND. Let $p_T$ be the payment of winners of class $T$. Then the critical value for all winning agents of class $T$ is $p_T$.*

For normalized, IC and ND mechanisms, we prove that the payment of winners of the same class are independent of their bids.

**Lemma 3.6.** *Let normalized $M$ be an IC and ND mechanism. For every agent class $T$, if agent $j$ of class $T$ wins and pays $p_T$ when he bids $b_j > p_T$ and a winning agent $i$ of class $T$ bids $b_i \geq p_T$, then $j$ also wins and pays $p_T$ if $i$ bids $\hat{b}_i > b_i$.*

*Proof.* Since $M$ is normalized and IC, then by Theorem 3.3 it is bid monotonic and by Observation 3.5 $p_T$ is the critical value for $i$ and $j$. By Observation 3.1 $p_T$ is independent of the bid $b_i$, so $i$ must pay $p_T$ if he bids $\hat{b}_i > b_i$. Since $M$ is non-discriminating, $j$ must also win (since $b_j > p_T$) and pay $p_T$. □

## 4. Incentive Compatible Double Auctions

Each of the markets along our supply chain performs a double auction (for a study of double auctions see Friedman & Rust, 1991). Our protocols for supply chains can work with a wide variety of double auction rules, and in this section we present several double auction rules that were suggested in the literature, and two new randomized double auction rules that we later use to create supply chain mechanisms. A double auction rule decides the trading agents and the payments to the agents.

We first give a description of the Double Auction (DA) model. It is a single-minded agents model with the following specifications of possible outcomes, value spaces and classes of agents. There is some homogeneous good $g$ that is traded in discrete quantities, and agents that are of one of two classes of agents. An agent is either a *seller* or a *buyer* of one unit of a good $g$. Seller $i$ has a single unit of $g$ and has some non-negative cost $s_i$ ($v_i = -s_i$) if he sells his unit. Buyer $i$ has a non-negative value $b_i$ ($v_i = b_i$) if he receives one unit of good $g$. The set of possible allocations are the set of *materially balanced* allocations, which are allocations with the same number of sellers and buyers.

A DA begins by agents reporting their values for one unit of the good. Each seller $i$ reports a cost $S_i$ which might be different from his real cost $s_i$. Each buyer $i$ reports a value $B_i$ which might be different from his real value $b_i$. All DA rules are non-discriminating by identity, so they first construct the *supply and demand curves* by sorting the supply bids s.t. $S_1 \leq S_2 \leq \ldots$ and the demand bids s.t. $B_1 \geq B_2 \geq \ldots$. If agents report truthfully then in order to maximize efficiency, once the trade quantity $q$ is set, the trading agents must be the first $q$ sellers and the first $q$ buyers. *The optimal trade quantity* $l$ is defined to be maximal such that $B_l \geq S_l$. Trading $l$ units maximizes the efficiency if the agents' bids are the agents' true values. In all double auctions we consider non-trading agents pay zero.

After setting the trade quantity to the optimal trade quantity, most real markets proceed by choosing a market clearing price anywhere in $[S_l, B_l]$. For example setting the price at $(B_l + S_l)/2$ – "the 1/2-Double Auction" which is a special case of **The k-Double Auction** (Wilson, 1985; Chatterjee & Samuelson, 1983; Satterthwaite & Williams, 1989). In general, the k-Double Auction is a double auction in which before the auction begins, a parameter $k$ is chosen such that $k \in [0, 1]$. $k$ is used to calculate a clearing price $P = k \cdot S_l + (1 - k) \cdot B_l$. $l$ units of the good are traded at the uniform price of $P$ for both sellers and buyers. This pricing scheme is *not* incentive compatible (in dominant strategies), but with additional strong assumptions on the agents values (all from a known distribution), it has been shown to perform reasonably well in many cases under strategic behavior of the participants (Rustichini, Satterthwaite, & Williams, 1994; Satterthwaite & Williams, 1991). Nevertheless, even when assuming a uniform distribution of the agents values, there is a non zero efficiency loss in the Bayesian Nash equilibrium achieved (and this equilibrium concept is much weaker than the dominant strategies equilibrium we discuss in this paper).

One may alternatively use **The VCG Double Auction** rule, which also sets the trade quantity to the optimal trade quantity $l$. This rule is non-discriminating, $l$ winning buyers pay $p_B = max(S_l, B_{l+1})$ and $l$ winning sellers receive $p_S = min(S_{l+1}, B_l)$. VCG mechanisms (Vickrey, 1961; Clarke, 1971; Groves, 1973) in general, and the VCG DA mechanism as a special case, are all IR and IC. Incentive compatibility leads to maximal efficiency since the trade is of the optimal size $l$, but it also leads to a budget deficit since $p_B \leq p_S$. By Observation 3.4 any normalized, IC and efficient mechanism must have the same payments as the VCG DA and thus the mechanism is not budget-balanced.

Myerson and Satterthwaite (1983) have shown that as long as individual rationality or "participation constraints" are met (e.g. if non-traders pay 0), an incentive compatible mechanism that always achieves the efficient outcome lead to a budget deficit (even for more general model and weaker solution concept that we consider). We now turn to look at several normalized and IC double auction rules that achieve budget balance (sometimes even surplus) at the price of achieving slightly sub-optimal efficiency (by reducing the trade quantity).

The simplest auction with this property is **The Trade Reduction (TR) DA**. In this non-discriminating auction, $l-1$ units of the good are traded. Each trading buyer pays $p_B = B_l$, and each trading seller receives $p_S = S_l$. Since $B_l \geq S_l$ this auction is BB. A rule which is an extension of the Trade Reduction DA rule, was previously suggested by McAfee (1992). By **McAfee's DA** rule, if a suggested clearing price $p = \frac{S_{l+1}+B_{l+1}}{2}$ is accepted by the $l$ buyer and seller ($p \in [S_l, B_l]$) then $l$ units of the good are traded at the price of $p$, otherwise the TR rule is used. Both the Trade Reduction DA rule and McAfee's rule are normalized, bid monotonic and the payments are by the critical values, thus they are IC (and IR) by applying Theorem 3.3. Both rules have efficiency of at least $(l-1)/l$ for any agents values, since only the unit with the lowest trade value might not be traded. McAfee's rule turned out not to be compatible with our protocols – we expand on this in appendix A.

We suggest two new normalized and randomized double auctions, which capture the tradeoff between the auction efficiency and the budget balance with one parameter $\alpha$. Unlike McAfee's rule, these rules can be used with our supply-chain protocols, and they achieve higher efficiency than the Trade Reduction DA in expectation. For a carefully chosen parameter $\alpha$ they also achieve exact budget balance (zero balance) on expectation (Ex ante Budget-Balanced).

In **The $\alpha$ Reduction DA**, for a fixed $0 \leq \alpha \leq 1$, the bids are submitted and then with probability $\alpha$ the Trade Reduction DA rule is used, and with probability $1-\alpha$ the VCG DA rule is used. This randomized double auction is *universally incentive compatible* (defined by Nisan & Ronen, 1999), which means that the agents bid truthfully even if they know the randomization result. This can be seen from the fact that bidding truthfully is a dominant strategy in both the Trade Reduction DA rule and the VCG DA rule. It is also normalized and ND since both Trade Reduction DA and VCG DA are normalized and ND.

**The $\alpha$ Payment DA** is another randomized double auction which has the same distribution of the allocation as the former auction (and therefore the same expected efficiency), but the payment of each agent is his expected payment of The $\alpha$ Reduction DA (so the expected budget of both auctions is the same). In this auction, a parameter $\alpha$ is chosen as in the $\alpha$ Reduction DA. Then the bids are submitted and the allocation and payments are decided. $l-1$ units are traded between buyers, each paying $\alpha \cdot B_l + (1-\alpha) \cdot max(B_{l+1}, S_l)$, and the sellers, each receiving $\alpha \cdot S_l + (1-\alpha) \cdot min(S_{l+1}, B_l)$. With probability $\alpha$ another unit of the good is traded between the $l$ buyer which pays $max(B_{l+1}, S_l)$ and the $l$ seller which receives $min(S_{l+1}, B_l)$. This auction is discriminating since the $l$ buyer pays a different price than the rest of the winning buyers if he wins the auction (and the same holds for the sellers). The $\alpha$ Payment DA has a much lower variance in most of the agents payments than the $\alpha$ Reduction DA, and it "hides" its randomized character from almost all the agents. The allocation and payment of all agents but the $l$ buyer and seller are independent of the coin toss, thus have zero variance. This auction is individually rational and incentive compatible; but not universally.

**Theorem 4.1.** *The $\alpha$ Payment DA is an incentive compatible randomized DA, but it is not a universally incentive compatible randomized DA.*

*Proof.* Since the $\alpha$ Reduction DA is universally incentive compatible, truth telling maximizes the expected utility of any agent in the $\alpha$ Reduction DA. The $\alpha$ Payment DA has the same probability distributions of the allocation and payments as the $\alpha$ Reduction DA, therefore truth telling maximizes the expected utility of any agent in the $\alpha$ Payment DA as well.

The $\alpha$ Payment DA is not a universally incentive compatible randomized DA. Assume that a seller which is one of the first $l-1$ sellers knows that the randomization will result in a VCG allocation, so he is now facing a deterministic auction. By Observation 3.4 there is a unique payment rule which ensures dominant strategy IC for normalized mechanism with the efficient allocation (the VCG allocation), this is the VCG payment rule. The $\alpha$ Payment DA is normalized and gives a different payment to that agent, therefore it cannot be incentive compatible in dominant strategies. □

Both auctions have efficiency of at least $(l-1)/l$, since at least $l-1$ of the $l$ units are traded. As $\alpha$ grows from zero to one, the expected revenue increases from negative to positive and the expected efficiency decreases (linearly). If the distribution $D$ of the agents values is known prior to the beginning of the auction, the parameter $\alpha = \alpha(D)$ can be chosen such that the expected revenue is zero (different distributions have different values of $\alpha(D)$, in this paper we do not solve the problem of finding $\alpha(D)$ analytically). We denote this value as $\alpha^*(D)$. For the uniform distribution of agents values in $[0,1]$ we denote this value as $a^*$.

In Table 1 we present a summary of the properties of the double auction rules.

| **DA rule** | **Incentive compatible** | **Revenue** | **efficiency loss** |
|---|---|---|---|
| k-DA | no | 0 | 0 |
| VCG | yes | deficit | 0 |
| Trade Reduction | yes | surplus | LFT |
| $\alpha^*(D)$ *Reduction* | yes, universally | 0 (expected) | $\alpha^*(D)$ *LFT* (expected) |
| $\alpha^*(D)$ *Payment* | yes, not universally | 0 (expected) | $\alpha^*(D)$ *LFT* (expected) |
| McAfee | yes | surplus | not more than LFT |

Table 1: Double Auction Rules comparison table

Notes : a) in the k-DA we assume that the agents bid truthfully, if they do not, there is a non zero efficiency loss in equilibrium. b) LFT means Least Favorable Trade which is $B_l - S_l$.

We have run simulations comparing the different DA rules with respect to the market revenue ($\sum_i p_i$) and to the total social efficiency ($\sum_{i \in A} v_i$). In Figures 5 and 6 we show the results for the average of 100 random executions of an auction with a given number of buyers and sellers, with all values drawn uniformly and independently at random in $[0, 1]$. We have calculated the $\alpha$ Reduction DA efficiency to revenue ratios for $\alpha \in \{0.25, 0.5, 0.75\}$. Note that in expectation, the same results would have been obtained for the $\alpha$ Payment DA, since it has the same expected efficiency and revenue as the $\alpha$ Reduction DA. We have calculated $a^* = \alpha^*(U)$ (where $U$ is the uniform distribution of agents values in $[0, 1]$) by simulations using binary search. Note that in both figures, the values of the Trade Reduction DA, the different $\alpha$ Reduction DA, the VCG DA as well as the $a^*$ Reduction DA, all lay on a linear curve. The point representing McAfee's DA lays above this line and in both simulations has lower efficiency than the $a^*$ Reduction DA. Our $a^*$ Reduction DA extracts more efficiency than McAfee's DA, while still being budget balanced. The simulations also show that in the symmetric case where the number of buyers and sellers is the same, all auctions extract a very high fraction of the efficiency (more than 99%). On the other hand in the asymmetric case where the number of buyers is much larger than the number of sellers, the Trade Reduction DA as well as McAfee's DA extract a significantly lower efficiency than the $a^*$ Reduction DA (about 5% less).

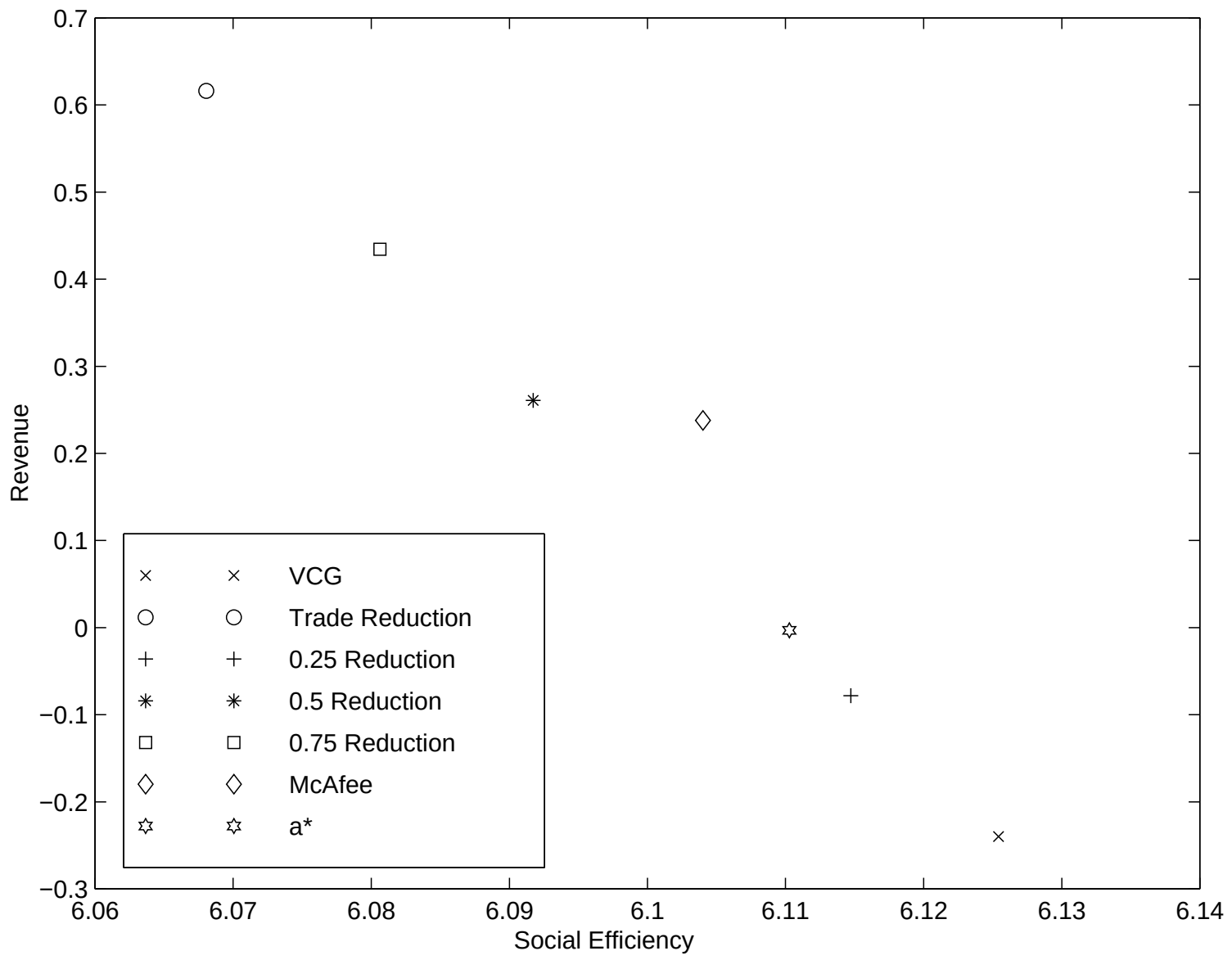

Figure 5: 25 buyers and 25 sellers
Revenue/Efficiency tradeoff simulations results

## 5. The Supply Chain Protocols

We begin by presenting the linear supply chain model, in which any unit of an initial good can be converted to a unit of the final good through a sequence of unit to unit conversions. Our supply chain model is a single-minded agents model with the following specifications of possible outcomes, value spaces and classes of agents. There is an ordered set $G$ ($|G| = t$) of homogeneous goods that are traded in discrete quantities. Each agent $i$ is of one of the following $t + 1$ classes. An *initial supplier* can supply one unit of the first good and has a

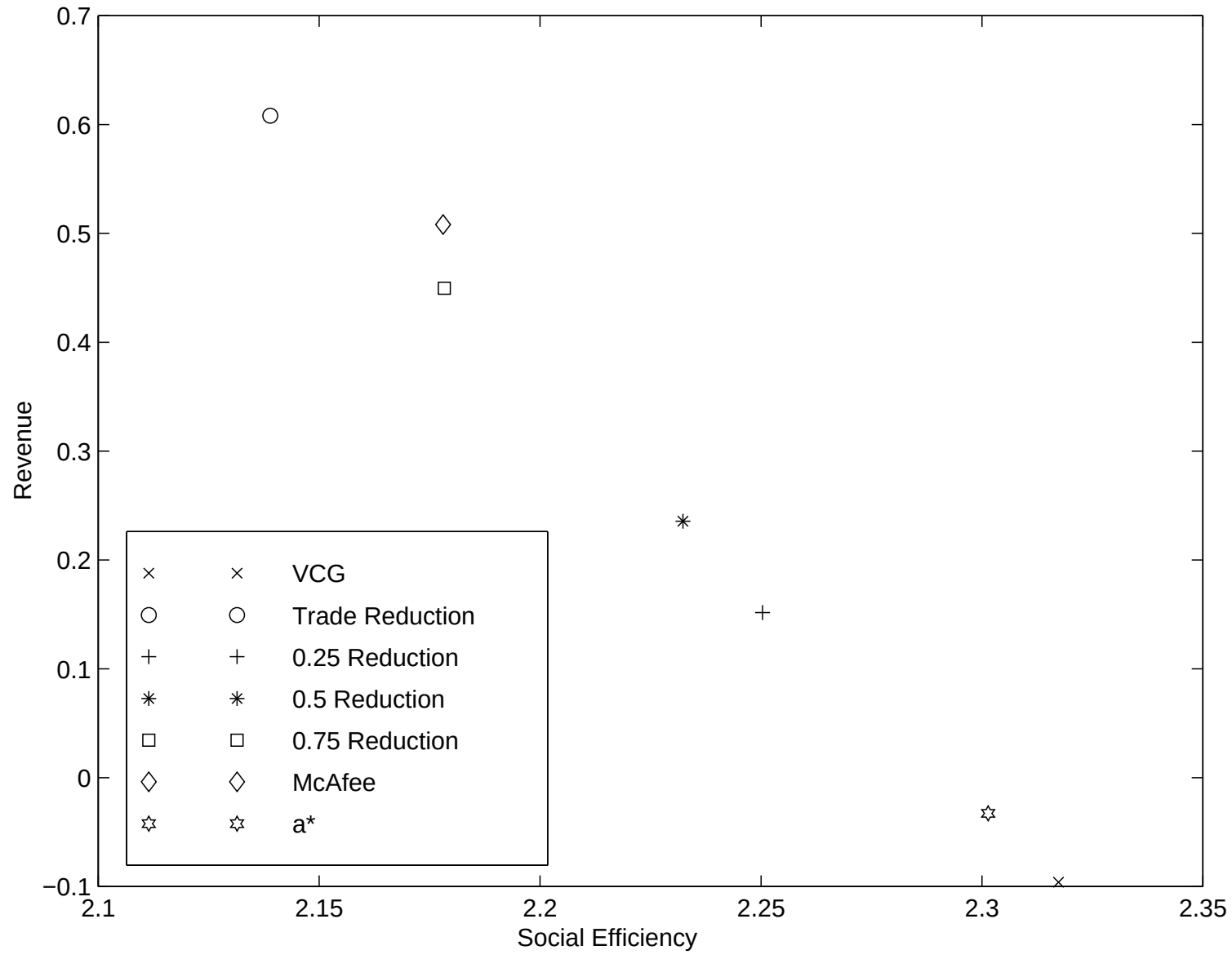

Figure 6: 50 buyers and 5 sellers
Revenue/Efficiency tradeoff simulations results non-negative cost $s_i$ ($v_i = -s_i$) if he trades. A *converter* of one unit of good $r \in \{1, \ldots, t-1\}$ to one unit of good $r+1$ has a non-negative cost $s_i$ ($v_i = -s_i$) if he converts a unit. A *consumer* has a non-negative value $b_i$ ($v_i = b_i$) if he receives one unit of the final good. A *market* is the set of agents of the same class. The allocation $A$ is *materially balanced* if for each good, the number of units produced is the same as the number of units consumed. This means that the number of winners of any class of agents is the same. Clearly the number of winners in any market cannot exceed the minimal number of bidders in any market $n$, so we can assume that all markets have only $n$ bidders ($n$ can be found by a single message between any market and the next market. The message includes the minimal size seen so far. The consumer market then sends $n$ backwards in the chain to all other markets).

We suggest two protocols that can be used to conduct an auction in a chain of markets in a distributed manner: **The Symmetric Protocol** and **The Pivot Protocol**. Both protocols run on servers connected by a network with a linear chain topology. Each server represents one market and receives bids from only one class of agents (suppliers, consumers or converters from some specified good to the next good in the chain). The servers do not act strategically, and follow the specified protocols (they are all owned by the same entity).

We denote the supply market as $M^1$ and each conversion market from a good $r$ to the following good $r+1$ as $M^{r \to r+1}$. The consumer (demand) market is marked by $M^t$. Each of the conversion markets is connected with bi-directional communication channels to the market that supplies its input good and to the market which demands its output good. We denote the supply and demand curves for a good $r$ as $S^r$ and $D^r$ respectively, and the supply and demand curves for the conversion of a good $r$ to a good $r+1$ as $S^{r \to r+1}$ and $D^{r \to r+1}$ respectively.

All agents must bid before some fixed deadline. The input to our protocols is the supply bids for the first good, the conversion bids for all the conversion markets and the demand bids for the final good. The protocols decide the allocation and the payments of the trading agents (the protocols are normalized, losing agents pay zero). Both our protocols are generic and can operate with various normalized DA rules.

In the Symmetric Protocol, each of the markets conducts a double auction, after constructing its demand and supply curves. Each agent views the supply chain auction as a double auction since each market conducts a double auction and there is no central market that makes the allocation and payments decisions. In order for the protocol to create a materially balanced allocation, some restriction are imposed on the double auction rule used by the Symmetric Protocol, as we later describe. This protocol can use either a discriminating or a non-discriminating DA rule. We present the protocol as using an abstract discriminating DA rule that takes supply and demand curves as inputs, and returns the trade quantity $q$ and two price vectors, $\vec{P}_S$ for the sellers and $\vec{P}_B$ for the buyers.

In the Pivot Protocol, only one market (the demand market) constructs its demand and supply curves, this market applies the double auction allocation and payments rule, and sends the results of the auction to it predecessor. Each market uses the information it receives and the bids in the market to send its predecessor information that are used to calculate its allocation and payments. Unlike the Symmetric Protocol, the Pivot Protocol must use a non-discriminating DA rule, but it creates a materially balanced allocation with any DA rule. For many DA rules, this protocol can be improved such that it has a much lower communication burden than the Symmetric Protocol, as we show in Section 5.3. We present the protocol as using an abstract non-discriminating double auction rule that takes supply and demand curves as inputs, and returns the trade quantity $q$ and a single price $P_S$ for the trading sellers and another price $P_B$ for the trading buyers.

### 5.1 The Symmetric Protocol

As we have seen in the example in Section 2, after the bids are submitted the Symmetric Protocol begins with supply curve propagation along the supply chain, from the supply market to the consumer market. The protocol continues with demand curve propagation along the supply chain in the other way (demand curves propagation may be done concurrently with the supply curves propagation). During this process each of the markets builds its supply and demand curves from the information it receives. At this point each of the markets has its supply and demand curves, and a double auction is conducted. If the rule is randomized, we assume that the random coins are public, which means that all markets have access to the same random coins (a public coin is created by one market tossing the coin and then propagating the result to all other markets along the supply chain). The formal protocol for the supply market $M^1$, the conversion markets $M^{r \to r+1}$ and the demand market $M^t$ are described in Figures 7, 8 and 9, respectively.

A DA rule decides on the trade size in each market according to the supply and demand curves created by the protocol. The allocation is materially balanced only in the case that the same trade size was decided in all the markets.

**Definition 4.** *Let $S(M)$ and $D(M)$ be the supply and demand curves created by the Symmetric Protocol in market $M$. Let $q(M)$ be the trade size obtained by applying the DA rule*

**Symmetric Protocol for the Supply Market $M^1$.**

1. $S^1 \leftarrow$ sort the list of supply bids in non-decreasing order.
2. Send $S^1$ to demand market $M^{1\to 2}$.
3. Receive $D^1$ from demand market $M^{1\to 2}$.
4. Apply the DA Rule on $(S^1, D^1)$ to obtain $(q, \vec{P}_S, \vec{P}_B)$.
5. Output: The $q$ lowest bidders sell a unit each and pay by the vector $\vec{P}_S$.

Figure 7: Symmetric Protocol for the Supply Market

**Symmetric Protocol for a Conversion Market $M^{r\to r+1}$**

1. $S^{r\to r+1} \leftarrow$ sort the list of supply bids in non-decreasing order.
2. When receiving $S^r$ from $M^{r-1\to r}$ ,
   send $S^{r+1} = S^r + S^{r\to r+1}$ to $M^{r+1\to r+2}$.
3. When receiving $D^{r+1}$ from $M^{r+1\to r+2}$,
   send $D^r = D^{r+1} - S^{r\to r+1}$ to $M^{r-1\to r}$.
4. Construct the market demand curve $D^{r\to r+1} = D^{r+1} - S^r$.
5. Apply the DA Rule on $(S^{r\to r+1}, D^{r\to r+1})$ to obtain $(q, \vec{P}_S, \vec{P}_B)$.
6. Output: The $q$ lowest bidders convert a unit each and pay by the vector $\vec{P}_S$.

Figure 8: Symmetric Protocol for a Conversion Market

**Symmetric Protocol for the Demand Market $M^t$**

1. $D^t \leftarrow$ sort the list of demand bids in non-increasing order.
2. Send $D^t$ to supply market $M^{t-1\to t}$.
3. Receive $S^t$ from supply market $M^{t-1\to t}$.
4. Apply the DA Rule on $(S^t, D^t)$ to obtain $(q, \vec{P}_S, \vec{P}_B)$.
5. Output: The $q$ highest bidders buy a unit each and pay by the vector $\vec{P}_B$.

Figure 9: Symmetric Protocol for the Demand Market

*$R$ on $(S(M), D(M))$. A DA rule $R$ is called* **consistent** *if $q(M1) = q(M2)$ for any two markets $M1$ and $M2$.*

The following Lemma shows that the optimal trade size in all markets is the same, so any rule that set the trade size as function of the optimal trade size is consistent.

**Lemma 5.1.** *Let $l(M)$ be the optimal trade size in market $M$. For any two markets $M1$ and $M2$, $l(M1) = l(M2)$. We denote this optimal trade size by $l$.*

*Proof.* By the definition of the optimal trade size, $l(M)$ is the maximal index such that $B_{l(M)}(M) \geq S_{l(M)}(M)$. It is enough to show that for any two markets $M1$ and $M2$, $B(M1) - S(M1) = B(M2) - S(M2)$ (as vectors).

For any $r$:

$D^{r \to r+1} - S^{r \to r+1} = (D^{r+1} - S^r) - S^{r \to r+1} = (D^t - \sum_{m=r+1}^{t-1} S^{m \to m+1}) - S^r - S^{r \to r+1} = D^t - (S^r + S^{r \to r+1} + \sum_{m=r+1}^{t-1} S^{m \to m+1}) = D^t - S^t$

Similar argument shows that $D^1 - S^1 = D^t - S^t$ □

We conclude that if the trade size $q_M$ decided by DA rule $R$ in any market $M$ is just a function of the optimal trade size $l_M = l$, then $R$ is consistent. Clearly a consistent DA rule always creates a materially balanced allocation.

Since the trade size decided by the VCG DA rule is $l$, then this rule is consistent. Similarly the Trade Reduction DA rule is consistent since the trade size is $l - 1$. Not all double auction rules are consistent, McAfee's DA is an example for an inconsistent rule (a specific example of inconsistency is presented in appendix A). The intuitive reason that it is not consistent is because the trade size is dependent on the comparison between some function of the $l + 1$ bids and the $l$ bids, which might be different between markets. On the other hand our two new randomized DA, the $\alpha$ Reduction DA and the $\alpha$ Payment DA are consistent. This is because all markets share a public coin, so either the trade size is $l - 1$ in all the markets, or it is $l$ in all the markets.

Following is the main theorem regarding the Symmetric Protocol.

**Theorem 5.2.** *Any normalized DA rule $R$ that is consistent can be used by the Symmetric Protocol to create a supply chain mechanism that is normalized and materially balanced. If $R$ is also IC then the mechanism created by the Symmetric Protocol is IC (thus also IR) and its efficiency is the efficiency of $R$. If $R$ is also non-discriminating, then the mechanism is also non-discriminating.*

*Proof.* Since $R$ is consistent the allocation is materially balanced by definition. The payments are normalized by definition, we show that if $R$ is IC then the mechanism is also IC. By Theorem 3.3 the DA rule $R$ is bid monotonic and the payments are by critical values. For any agent, the supply and demand curves (disregarding his bid) are independent of his bid by the way they are built by the protocol. So the mechanism is bid monotonic and the payments are by critical values (from the agent point of view, he submits his bid to a DA), therefore by the other direction of Theorem 3.3 the supply chain mechanism is IC (and IR). By IC and the way that costs are aggregated by the protocol, the efficiency of the mechanism is the same as the efficiency of $R$. If the $R$ is non-discriminating, then the payments of all the winning agents in the any market are the same, so the mechanism is non-discriminating. □

Note that the same normalized, IC and consistent DA rule must be used in all the markets for the mechanism to be materially balanced, normalized and IC. The DA rule is just a function of the supply and demand curves in the market. If any two of the supplier and conversion markets get the same supply bids, they also get the same demand curves so for consistency both must have the same allocation. Since both are normalized and IC by Observation 3.4 both must also have the same payment rule. The demand market must have the same DA rule as the supply market, since in the case that all converters have a zero cost, both markets have the same supply and demand curves. Applying the same argument presented above to these markets, proves that they must have the same DA rule.

## 5.2 The Pivot Protocol

The Pivot Protocol is a supply chain protocol that creates a normalized materially balanced supply chain mechanism from any non-discriminating normalized double auction rule. If the DA rule is also incentive compatible then the protocol creates an incentive compatible supply chain mechanism. Unlike the Symmetric Protocol, the Pivot Protocol is not restricted to using consistent double auction rules, so in that sense it is less restricted than the Symmetric Protocol. On the other hand, unlike the Symmetric Protocol it is restricted to use non-discriminating double auction rules. The use of randomized double auctions is also more natural in the Pivot Protocol, since only one market (the Pivot market) uses the random coins and there is no need for public coins created by distribution of the random coins of one market to all other markets.

Before defining the Pivot Protocol formally, we explain its execution on the Lemonade-stand industry example of Section 2. As in the Symmetric Protocol, the protocol begins with supply curve propagation along the supply chain to the juice market (see Figure 2). Figure 10 presents the information propagation of the Pivot Protocol after the supply curve propagation stage, and is explained below.

| **Lemon Market** | |
|---|---|
| **Supply** ($S^L$) | **Demand** ($D^L$) |
| **3** | |
| **6** | 8 |
| **7** | |

| **Squeezing Market** | |
|---|---|
| **Supply** ($S^{L\to J}$) | **Demand** ($D^{L\to J}$) |
| **1** | |
| **3** | 5 |
| **6** | |

| **Juice Market** | |
|---|---|
| **Supply** ($S^J$) | **Demand** ($D^J$) |
| 4 | **12** |
| 9 | **11** |
| 13 | **7** |

(V,q)=(8,2) ←

(V,q)=(11,2) ←

Figure 10: The Pivot Protocol for the Lemonade-stand industry

Unlike in the Symmetric Protocol, the demand curve is not propagated backwards, but rather, the juice market, which has its supply and demand curves at that stage, now conducts a non-discriminating double auction. For example, running the VCG double auction results with efficient trade of size 2 in the juice market. This trade size is propagated to all other markets and is set to be the trade size in each market along the supply chain, ensuring a materially balanced allocation (the trade size is denoted by $q$ in Figure 10). The VCG DA rule charges the two winning consumers in the juice market for $\$9 = max(\$9, \$7)$ each (since they must bid at least \$9 to match the supply of juice, and if they do, they also defeat Cindy, the losing consumer). There is no single agent that has a juice supplier role in the juice market, rather, a "juice supplier" is an aggregation of a lemon picker and a juice squeezer. We use the price that the double auction would have charged a supplier in the juice market to find the prices for lemon pickers and juice squeezers. Each market sends to it predecessor in the chain the highest price the winning agents are willing to pay for one unit of the input good. A "supplier" in the juice market should be paid $\$11 = min(\$13, \$11)$, the maximal cost a winning supplier can charge for juice. So, \$11 is the highest cost for juice, and the juice market informs the squeezing market that the trade size is 2 and that the price of juice should not exceed \$11 ($V$ in the figure marks this propagated value). The squeezing market informs the lemon market that the trade size is 2 and that the price in the lemon market should be at most $\$8 = \$11 - \$3$, since the highest cost of a winner in the squeezing market is \$3. The price in the squeezing market cannot exceed $\$5 = \$11 - \$6$, since the price of juice cannot exceed \$11 and the cost of lemon for a winning lemon picker might be as high as \$6. Also, the price in the squeezing market cannot exceed \$6, the cost of the loser in that market, so the price is set to $\$5 = min(\$6, \$5)$. The price in the lemon market is set to $\$7 = min(\$7, \$8)$, since the price cannot exceed \$7, the cost of the loser in that market, which is lower than the propagated maximal cost of \$8.

Note that when we use the VCG double auction rule, the Lemonade-stand industry allocation and payments of the Pivot Protocol are exactly the same as the allocation and payments of the Symmetric Protocol presented in Section 2. We later show (Theorem 5.5) that this is a general phenomena which happens for any normalized DA incentive compatible rule that can be used by both protocols, that is, a DA rule that is non-discriminating and consistent (if the DA rule is inconsistent, like the McAfee's DA rule, the allocation created by the Symmetric Protocol is not materially balanced. If on the other hand the rule is discriminating, it cannot be used by the Pivot Protocol since there are different maximal unit costs for different winning agents in the same market).

We now turn to the general definition of the Pivot Protocol. In the Pivot Protocol, one of the markets is chosen as a pivot and the double auction is only held there. Any market may be chosen as a pivot; we describe the case where the pivot is the consumer market. After all bids are submitted, the protocol begins with supply curves propagation along the supply chain as in the Symmetric Protocol. At this point the consumer market has its supply and demand curves, and a non-discriminating double auction is conducted in this market. The DA sets the trade size, the price the mechanism charges the consumers, and the highest cost of a unit of the good (supplier's price), which is sent to the predecessor market of the consumer market. After that, starting with the consumer market which is now viewed as the demand market of its predecessor, each of the demand markets sends to its supply market the size of trade and the highest price that the demand market is willing to pay for one unit of its input good, without reducing the trade quantity. The payment in each market is set to the maximal cost in each market that results with the same trade size, and is calculated as the minimum of two terms. The first term is the difference between the propagated maximal cost of the market output and the maximal cost of the market input. The second term is the highest cost of a loser in that market. We show below that this payment scheme creates a normalized, incentive compatible and non-discriminating supply

**Pivot Protocol for Pivot Market** $M^t$

1. $D^t \leftarrow$ sort the list of demand bids in non-increasing order.
2. Receive $S^t$ from supply market $M^{t-1 \to t}$.
3. apply DA Rule on $(S^t, D^t)$ to obtain $(q, P_S, P_B)$.

// Send results to the other markets:

4. Send $(P_S, q)$ to market $M^{t-1 \to t}$.
5. Output: The $q$ highest bidders buy for price $P_B$.

Figure 11: The Pivot Protocol for Pivot Market

**Pivot Protocol for a Conversion Market** $M^{r \to r+1}$

1. $S^{r \to r+1} \leftarrow$ sort the list of supply bids in non-decreasing order.
2. When receiving $S^r$ from $M^{r-1 \to r}$ ,
   send $S^{r+1} = S^r + S^{r \to r+1}$ to $M^{r+1 \to r+2}$.
3. When receiving the pair $(V, q)$ from $M^{r+1 \to r+2}$,
   send $(V - S_q^{r \to r+1}, q)$ to $M^{r-1 \to r}$.
4. Output: The $q$ lowest bidders sell for price $min(V - S_q^r, S_{q+1}^{r \to r+1})$.

Figure 12: Pivot Protocol for a Conversion Market

chain mechanism, given a normalized, incentive compatible and non-discriminating double auction rule.

The formal protocol for the pivot market, the conversion markets and the supply market are presented in Figures 11, 12 and 13, respectively.

The following theorem shows how the properties of the mechanism created by the Pivot Protocol are derived from the properties of the DA rule used by the pivot market.

**Theorem 5.3.** *Any normalized DA rule R that is non-discriminating can be used by the Pivot Protocol to create a normalized supply chain mechanism that is materially balanced and non-discriminating. If R is also IC then the mechanism created by the Pivot Protocol is IC (thus also IR) and its efficiency is the efficiency of R.*

**Pivot Protocol for the Supply Market** $M^1$

1. $S^1 \leftarrow$ sort the list of supply bids in non-decreasing order.
2. Send $S^1$ to demand market $M^{1 \to 2}$.
3. Receive the pair $(V, q)$ from the demand market $M^{1 \to 2}$.
4. Output: The $q$ lowest bidders sell for price $min(V, S_{q+1}^1)$.

Figure 13: Pivot Protocol for the Supply Market

*Proof.* From the protocol definition the mechanism is materially balanced, normalized and non-discriminating. Assume that the DA rule is normalized and IC. Then, since by the protocol definition the supply curve in the pivot market is independent of the bids of the consumers, the mechanism is normalized and IC for the consumers. We prove that it is IC for initial suppliers, the proof for the converters is similar (so it will be omitted), and we conclude that the mechanism is normalized and IC (thus also IR). We show that the mechanism is bid monotonic and the payments are by critical values, thus by Theorem 3.3 the mechanism is IC.

Assume supplier $i$ bids $S_i^1$ and wins. Assume the trade size is $q$ ($q \geq i$ since $i$ wins) and $P_S$ is the price to the "sellers" in the pivot market. The allocation is bid monotonic since if he bids a lower cost he is still one of the $q$ lowest bidding suppliers and his bid change can only reduce the cost of the $q$ lowest bidding sellers in the pivot market. By applying Lemma 3.6 on the DA rule, $P_S$ remains the same, and $i$'s corresponding seller bid is still lower than $P_S$ so he remains a winner. Let $C_i^1$ be the critical value for supplier $i$.

By the Pivot Protocol the payment to a winning supplier in the supply market is

$$p_i^1 = min(P_S - \sum_{r=1}^{t-1} S_q^{r \to r+1}, S_{q+1}^1)$$

We need to show that $C_i^1 = p_i^1$ to prove that the payments are by critical values. Assume that $i$ changes his bid to $X$ and he is now the $j$ bid in his market under the new order of the bids. We show that if $X < p_i^1$ he wins and if $X > p_i^1$ he loses (note that here we consider costs, not values). This means that $p_i^1$ is the critical value for supplier $S_i^1$ to win the auction.

- If $X < p_i^1$ then $X < S_{q+1}^1$ and the supplier is still one of the $q$ highest bidders in his market ($j \leq q$). By the way the protocol builds the supply curve in the pivot market, his bid change can only change the $q$ highest sellers bids in the pivot market. All the $q$ highest sellers bids in the pivot market are still below $P_S$, since originally $S_q^1 + \sum_{r=1}^{t-1} S_q^{r \to r+1} < P_S$, and even if he is the $q$ bidder by the assumption $X + \sum_{r=1}^{t-1} S_q^{r \to r+1} < P_S$. By Lemma 3.6 applied on the DA rule, $P_S$ remains the same and since $i$'s corresponding seller bid is below $P_S$, the supplier wins the auction. This proves that $C_i^1 \geq p_i^1$

- If $X > S_{q+1}^1$, then $j \geq q+1$. For him to win when bidding $X$ it must be that $C_i^1 > X + \sum_{r=1}^{t-1} S_j^{r \to r+1}$, but then $C_i^1 > S_{q+1}^1 + \sum_{r=1}^{t-1} S_{q+1}^{r \to r+1}$ which by ND means that the original $q+1$ supplier (which is now the $q$ supplier) must win if $i$ bids $X$. By Lemma 3.6 the original $q+1$ supplier must also win if $i$ bids $S_i^1$ which contradicts the assumption that the trade size is $q$ when $i$ bids $S_i^1$.

- If $X > P_S - \sum_{r=1}^{t-1} S_q^{r \to r+1}$ and $X \leq S_{q+1}^1$ then $j \leq q$ (without loss of generality he is ordered before the $q+1$ bid if there is a tie). Since $X + \sum_{r=1}^{t-1} S_q^{r \to r+1} > P_S$ and $P_S \geq S_q^1 + \sum_{r=1}^{t-1} S_q^{r \to r+1}$ (the original $q$ supplier wins when $i$ bids $S_i^1$) we conclude that $X > S_q^1$ which means that he is now the $q$ bidder ($j = q$). So his corresponding seller bid in the DA is $X + \sum_{r=1}^{t-1} S_q^{r \to r+1} > P_S$, which is not high enough to win the DA and therefore the agent loses.

We conclude that any agent which wins the auction pays his critical value, and therefore the auction is IR and IC. By IC and the way that costs are aggregated by the protocol, the efficiency of the mechanism is the same as the efficiency of $R$. □

Theorem 5.3 has presented the relationship between incentive compatibility of the DA rule and the resulting supply chain auction created by the Pivot Protocol. This brings up the question of whether the budget balance of the DA rule ensures budget balance of the supply chain mechanism. It turn out not to be so! In appendix A we present an example of a DA rule (McAfee's rule) which has a revenue surplus, but the Pivot Protocol using that rule creates a mechanism with a revenue deficit.

The next theorem gives a condition on the DA rule which is sufficient to ensure that the supply chain mechanism created by the Pivot Protocol is budget-balanced. It also shows that a DA rule with budget deficit creates a mechanism with a deficit.

**Theorem 5.4.** *Let $R$ be a normalized DA rule that is IC and non-discriminating. Let $M$ be the supply chain mechanism created by the Pivot Protocol using the rule $R$ . For any supply and demand curves, let $q$ be the trade size decided by $R$, and let $P_S$ and $P_B$ be the sellers and buyers prices respectively.*

*If $P_B \geq S_{q+1}$ holds for all supply and demand curves, then $M$ is budget-balanced. If $P_B < P_S$ for some supply and demand curves ($R$ has budget deficit) then $M$ has budget deficit.*

*Proof.* We first show that if $P_B \geq S_{q+1}$ holds for all supply and demand curves, then $M$ is budget-balanced. We divide the supply chain allocation to $q$ disjoint procurement sets. Each set has a single winner from each market. We show that the total payment from each procurement set is non-negative, therefore the auction is BB. By Theorem 5.3 the mechanism is non-discriminating, so all procurement sets have the same payments. Let $P$ be the total payments to all agents in a procurement set, except for the consumer. Lemma B.1 (see appendix B) shows that $S^t_{q+1} \geq P \geq P_S$. If $P_B \geq S^t_{q+1}$ then $P_B \geq P$, and we conclude that $P_B - P \geq 0$ and $M$ is BB.

Similarly if for some supply and demand curves $R$ has budget deficit, then every procurement set has a deficit. If $P_B < P_S$ for some supply and demand curves, then $P_B < P$ and we conclude that the total payment for each procurement set is $P_B - P < 0$, which means that $M$ has a budget deficit. □

Note that for a double auction rule $R$ to be BB, it is sufficient that $P_B \geq P_S$. To prove that the supply chain mechanism is budget balanced, the theorem relies on a stronger condition that $P_B \geq S_{q+1}$ (it is always true that $S_{q+1} \geq P_S$ as Lemma B.1 shows). While McAfee's DA satisfies $P_B \geq P_S$, the stronger condition $P_B \geq S_{q+1}$ does not always hold, and indeed the example in Appendix A shows that the supply chain mechanism has a budget deficit in some cases. On the other hand, the Trade Reduction DA rule satisfies both constraints since $q = l - 1$ (recall that $l$ is the optimal trade size) and $P_B = B_{q+1} = B_l \geq S_l = S_{q+1} = P_S$.

Clearly a DA rule that has budget deficit might create a supply chain mechanism with budget deficit if used by the Symmetric Protocol. This is since double auction is a special case of a supply chain. We were not able to present a parallel sufficient condition for the Symmetric Protocol to ensure a budget-balanced supply chain mechanism. For a case that a normalized DA rule is incentive compatible, consistent and non-discriminating, the next theorem proves that the two protocols create the same mechanism, which implies that the above sufficient condition for budget balance holds also for the Symmetric Protocol with such DA rules.

We now turn to look at the relationship between the two protocols. In case that the normalized DA rule is consistent and non-discriminating it can be used by both protocols. The following theorem shows that if the rule is also IC, then the Symmetric Protocol and the Pivot Protocol create the same mechanism:

**Theorem 5.5.** *Let $R$ be a normalized DA rule that is non-discriminating and consistent. If $R$ is IC then the mechanism created by the Pivot Protocol using $R$ is the same as the mechanism created by the Symmetric Protocol using $R$ (for any set of bids, both have the same allocation and payments). Thus, the efficiency and the budget of both mechanisms are equal.*

*Proof.* Since the supply and demand curves of the demand market $M^t$ are built in the same way, and the DA rule used by the demand market is the same in both protocols, the market trade sizes decided by the rule are the same in both mechanisms. The Symmetric Protocol is consistent, therefore the trade sizes in all the markets are the same. Since the mechanism is non-discriminating, the allocations are the same in all the markets in both protocols. By Observation 3.4 both mechanisms must have the same payments, since they have the same allocation. □

### 5.3 Communication Complexity

A naive implementation of a supply chain mechanism as a centralized mechanism would have required sending $\Omega(tn)$ bids to one centralized point ($t$ is the number of goods and $n$ is the minimal number of agents in any market). Both protocols that we have described have much lower communication of $O(n)$ prices received and sent to each market (note that if a bid of a single agent requires $k$ bits, then the prices communicated do not grow too much and have at most $k + t$ bits). For many interesting DA rules we can further reduce the communication of each market to only $O(log(n))$ prices by the improved Pivot Protocol we present below. Thus, for these DA rules the Pivot Protocol can be implemented for exponentially larger markets when still using the same bandwidth. For example, we can use the improved protocol if the size of trade $q$ set by the DA rule can be calculated from the Optimal Trade Quantity $l$, and the payments are dependent on the $l$ and $l + 1$ bids in the supply and demand curves only (this is the case for all DA rules we have presented).

**Theorem 5.6.** *Let $R$ be a normalized DA rule that is IC and non-discriminating. Assume that the trade size decided by this rule is a function of the Optimal Trade Size $l$ only, and the payments can be decided using $O(1)$ prices when knowing $l$. Then the Pivot Protocol can be implemented with only $O(log(n))$ messages sent and received by each market, where each message contains a single price.*

*Proof.* The Pivot Protocol can be improved by using binary search to find $l$ and thus sending only the few values needed from the supply curves, instead of passing entire supply curves along the chain. The search for $l$ can be preformed by a binary search while sending only $O(log(n))$ messages between any two consecutive markets. In the improved protocol, first $S^t_{\frac{n}{2}}$ is passed to the pivot market, for $n$ which is the minimal number of bids in any market (known to all markets by the protocol explained in the beginning of this section). Then the pivot market checks if this value is smaller or greater than $D^t_{\frac{n}{2}}$ and asks for the $\frac{3 \cdot n}{4}$ or the $\frac{n}{4}$ element of the supply curves respectively. The pivot market receives the requested value and continues in a similar way with the search, until $l$ is found. Then $O(1)$ prices needed from the supply curve to find the payments (which are only a function of $l$ by our assumption) are propagated to the pivot market by its request. □

Similar results can be presented for the Symmetric Protocol with the same restrictions on the DA rule used by the protocol. The Symmetric Protocol is the protocol of choice when the DA rule is discriminating, but in this case it is more likely that entire supply and demand curves are needed in order to find the prices for all winners.

### 5.4 Global Properties of the Different DA Types

In this section we examine the properties of the mechanisms created by the supply chain protocols using different double auction rules.

**Chain of VCG Double Auctions** is created by running the Symmetric Protocol using the VCG DA rule. This creates a local VCG double auction in each of the markets. Since VCG DA rule is IR, IC, consistent and non-discriminating, by Theorem 5.5 the Pivot Protocol with the VCG DA rule creates the same mechanism.

The following proposition summarizes the properties of the supply chain mechanism using the VCG double auction rule.

**Proposition 5.7.** *Chain of VCG Double Auctions is IR, IC, non-discriminating, Socially Efficient and has a revenue deficit.* $(R_{Chain\ VCG} \leq 0)$

*Proof.* Since VCG DA rule is normalized, IC and non-discriminating, by Theorem 5.3 the mechanism created is IR, IC and non-discriminating. The outcome maximizes the sum of valuations (Socially Efficient), since agents bid truthfully, the trade size is the Optimal Trade Size and the highest value bidders in each of the markets are chosen as winners.

By the fact that VCG DA has budget deficit and Theorem 5.4 we conclude that the Chain of VCG DAs has budget deficit. □

Note that by Observation 3.4 the mechanism created using VCG DA rule is VCG in the global sense presented by Vickrey, Clarke and Groves (A mechanism that is IR, IC and efficient).

**Chain of Trade Reduction Double Auctions** is created by running the Symmetric Protocol using the Trade Reduction DA Rule. This creates a local Trade Reduction double auction in each of the markets, and those double auctions are chained. Since the Trade Reduction DA rule is normalized, IC, consistent and non-discriminating, by Theorem 5.5 the Pivot Protocol with the Trade Reduction DA rule creates the same mechanism.

**Proposition 5.8.** *Chain of Trade Reduction Double Auctions is IR, IC , non-discriminating, has a revenue surplus of*

$$R_{Chain\ Trade\ Reduction} = (l-1) \cdot (D^t_l - S^t_l) \geq 0$$

*the reduction in the Social Efficiency is $D_l^t - S_l^t$ and its efficiency is at least $(l-1)/l$ of the maximal efficiency.*

*Proof.* Since the Trade Reduction DA rule is normalized, IC and ND, by Theorem 5.3 the mechanism created is IR, IC and non-discriminating.

The revenue is the sum of the payments which is

$$(l-1) \cdot (D_l^t - \sum_{r=1}^{t-1} S_l^{r \to r+1} - S_l^1) = (l-1) \cdot (D_l^t - S_l^t)$$

and it is non-negative since $D_l^t \geq S_l^t$ by the definition of The Optimal Trade Quantity $l$.

The reduction in social efficiency is $D_l^t - S_l^t$, since the $l$ unit is not traded, so as in the Trade Reduction DA the $l-1$ highest trades out of the optimal $l$ trades are conducted, giving efficiency of at least $(l-1)/l$. □

**Chain of $\alpha$ Reduction Double Auctions** is created by running the Pivot Protocol using the $\alpha$ Reduction Rule, which creates a randomized mechanism. This mechanism is a probability distribution over two mechanisms - the Chain of Trade Reduction DA and the Chain of VCG DA. The mechanisms are used according to a random variable $\chi$ that is one with probability $\alpha$ and zero otherwise. The first mechanism is chosen if $\chi$ is one (with probability $\alpha$), and the second mechanism is chosen if $\chi$ is zero (which happens with probability $1-\alpha$).

This auction can also be preformed using the Symmetric Protocol under the public coin assumption (which can be created by distributing a random coin, tossed by one of the markets). In this case the random variable $\chi$ is shared by all the markets, meaning that either the Trade Reduction rule is used by all the markets, or by non of the markets. This ensures trade consistency and creates the same mechanism.

Chain of $\alpha$ Reduction Double Auctions achieves the allocation and payments of the Chain of Trade Reduction DA with probability $\alpha$, and with probability $1-\alpha$ it achieves the allocation and payments of the Chain of VCG DA.

**Proposition 5.9.** *Chain of $\alpha$ Reduction Double Auctions is an individually rational and universally incentive compatible randomized mechanism. Its expected revenue is*

$$R_{Chain\ \alpha\ Reduction} = \alpha \cdot R_{Chain\ Trade\ Reduction} + (1-\alpha) \cdot R_{Chain\ VCG}$$

*its efficiency is at least $(l-1)/l$ of the maximal efficiency. Its expected reduction in the Social Efficiency is $\alpha \cdot (D_l^t - S_l^t)$ when the expectation is over the random choice of $\chi$.*

*Proof.* The mechanism is individually rational and universally incentive compatible randomized mechanism, since both the Chain of Trade Reduction DAs and the Chain of VCG DAs are normalized and incentive compatible by Proposition 5.8 and Proposition 5.7.

It is easy to verify that the expected revenue and the reduction in social efficiency are as claimed. □

**Chain of $\alpha$ Payment Double Auctions** is created by running the Symmetric Protocol, and in each market running the $\alpha$ Payment Double Auction, which creates a randomized mechanism (again under the public coin assumption). The mechanism has the same expected revenue and social efficiency as The Chain of $\alpha$ Reduction DAs, and each of the agents has the same probability of winning the auction and the same expected utility as in The Chain of $\alpha$ Reduction DAs. Note that since the $\alpha$ Payment DA rule is discriminating, the Pivot Protocol as defined above cannot be used to create a supply chain mechanism.

**Proposition 5.10.** *Chain of $\alpha$ Payment Double Auctions is an individually rational and incentive compatible randomized mechanism. Its expected revenue and its expected social efficiency is the same as The Chain of $\alpha$ Reduction Double Auctions.*

*Proof.* The Chain of $\alpha$ Payment Double Auctions has the same distribution of the allocation and payments as The Chain of $\alpha$ Reduction Double Auctions, therefore each agent is facing the same mechanism in expectation so he bids truthfully. Since the expected payment to each agent is the same in both mechanisms, the expected revenue is the same. Since the allocation is the same in both mechanisms, the expected social efficiency is the same. □

Unfortunately, chaining McAfee's Double Auction does not preserve the nice properties McAfee's DA rule has. If we run the Symmetric Protocol using McAfee's DA Rule, we might get an inconsistent trade. On the other hand, if we run the Pivot Protocol using McAfee's DA rule, we have a revenue deficit! (See examples in appendix A). However a weaker claim can be made - this mechanism has revenue at least as high as the chain of VCG DA, and has efficiency at least as high as the chain of Trade Reduction DA.

**Chain of McAfee's Double Auction** is created by running the Pivot Protocol using McAfee's Rule.

**Proposition 5.11.** *Chain of McAfee's Double Auction is IR, IC, ND, its revenue is never smaller than the revenue of the Chain of VCG Double Auction, and its reduction in the Social Efficiency is never greater than the reduction of the Chain of Trade Reduction Double Auction.*

*Proof.* Chain of McAfee's Double Auction is IR, IC and ND by Theorem 5.3 and the fact that McAfee's DA rule is normalized, IC and ND.

The mechanism revenue is never smaller than the revenue of the Chain of VCG Double Auctions since if there is trade reduction, then there is a revenue surplus by Proposition 5.8, and we know that Chain of VCG Double Auctions has a revenue deficit by Proposition 5.7. In the case of no trade reduction, the revenue is never smaller than the revenue of the Chain of VCG Double Auctions, since each winning buyer pays $p = P_B = P_S = \frac{S^t_{l+1}+D^t_{l+1}}{2}$ which is at least as much as he pays in the VCG mechanism. This is because $S^t_{l+1} > D^t_{l+1}$ so $S^t_{l+1} > p > D^t_{l+1}$ and $S^t_l \leq p$ and therefore $max(D^t_{l+1}, S^t_l) \leq p$. Each winning seller of the supply good, never receives more than $min(S^1_{l+1}, D^1_l)$ which he receives in the VCG mechanism. This is since $p \leq min(S^t_{l+1}, D^t_l)$ and he receives

$$min(S^1_{l+1}, p - \sum_{r=1}^{t-1} S_l^{r\to r+1}) \leq min(S^1_{l+1}, min(D^t_l, S^t_{l+1}) - \sum_{r=1}^{t-1} S_l^{r\to r+1}) =$$

$$min(S^1_{l+1}, D^t_l - \sum_{r=1}^{t-1} S_l^{r\to r+1}, S^t_{l+1} - \sum_{r=1}^{t-1} S_l^{r\to r+1}) = min(S^1_{l+1}, D^1_l)$$

Where the last equality holds since $D_l^1 = D_l^t - \sum_{r=1}^{t-1} S_l^{r\to r+1}$ and $S_{l+1}^t - \sum_{r=1}^{t-1} S_l^{r\to r+1} \geq S_{l+1}^t - \sum_{r=1}^{t-1} S_{l+1}^{r\to r+1} = S_{l+1}^1$. A similar argument shows that each winning convert never receives more than in the VCG mechanism.

The reduction in the social efficiency is never greater than the reduction in the Chain of Trade Reduction Double Auction, since the trade size is never smaller than the trade size in this auction, which is the $l-1$. □

Table 2 summarizes the properties of the supply chain mechanisms created by the two protocols using different double auction rules. For normalized mechanisms, if one cares about incentive compatibility and efficiency but not about budget balance, then the VCG mechanism should be used. If budget balance must be ensured for **every** execution of an incentive compatible mechanism, then the Trade Reduction auction is the mechanism of choice. If one wants the mechanism to be incentive compatible for every execution, but can settle for expected budget balance instead of ensured one, then higher expected efficiency can be achieved by the $\alpha^*(D)$ *Reduction*, assuming the right choice of $\alpha^*(D)$. Finally, if the variance in the agents payments is important, then the $\alpha^*(D)$ *Payment* should be used in order to lower the variance for most agents. This has a cost, since now the mechanism is not incentive compatible for every execution (universally), but rather, it is incentive compatible when the agents only care about expected utility. The table also presents the McAfee supply chain mechanism, to point out the fact that the budget of the supply chain mechanism might be different from the budget of the underlying double auction.

| **DA rule** | **Incentive Comp.** | **Revenue** | **Efficiency loss** |
|---|---|---|---|
| VCG | yes | deficit | 0 |
| Trade Red. | yes | surplus | LFT |
| $\alpha^*(D)$ *Reduction* | yes, universally | 0 (expected) | $\alpha^*(D)$ *LFT* (expected) |
| $\alpha^*(D)$ *Payment* | yes, not universally | 0 (expected) | $\alpha^*(D)$ *LFT* (expected) |
| McAfee | yes | surplus or deficit | not more than LFT |

Table 2: Supply Chain Auctions

Notes: a) McAfee's rule can not be used by the Symmetric Protocol, the table presents its properties under the Pivot Protocol. b) LFT (least favorable trade) in the context of supply chain, means the net total utility of the least favorable item. c) The distribution $D$ is over all agents in the supply chain. d) $\alpha^*(D)$ *Payment* has a lower variance in the payments than the $\alpha^*(D)$ *Reduction*.

When comparing the properties of the chain mechanisms in Table 2 with the properties of the original double auction rules in Figure 1, one can see the following. The incentive compatibility and the efficiency properties are preserved by the protocols, while achieving a revenue surplus requires a stronger condition on the DA rule, as we have shown in Theorem 5.4. The consistency property, which enables chaining of the markets by the Symmetric Protocol, applies to the first two deterministic rules, and the two randomized rules under the assumption of common coin toss. It does not exist in McAfee's rule since this rule sets its trade quantity as a function of the bids submitted, in such a way that the trade quantity can be different in two different markets. The k-DA is not presented in the table since it is not incentive compatible and therefore it cannot be used by the protocols (the two protocols create different mechanisms with this rule).

## 6. Discussion and Future Work

In this paper we have presented two distributed protocols that create supply chain mechanisms using double auction rules. Our protocols are generic and can use different double auction rules to create different mechanisms. We have characterized the properties of the supply chain mechanisms which derive from the properties of the underlining double auction rule. The protocols can use the VCG DA to achieve IR, IC and full efficiency with budget deficit. They can also use the Trade Reduction DA to ensure budget balance or surplus with high efficiency. An even higher efficiency while maintaining BB in expectation can be achieved using our two new randomized double auction rules.

Our work concentrated on the simple case of indivisible homogeneous goods, where each of the agents desires only one good or desires to convert one unit of good to one unit of another good, and the supply chain has a linear form. Our Pivot Protocol is easily extended to the case of trees in which each good can be created from one of several goods (but no good is used to create two different goods). For example, we might have a lemon market in Florida and another lemon market in California, and also separate squeezing markets in both states, but only one lemonade market. The tree of auctions decides how many lemonade glasses will be produced, and how many of the glasses will be produced from Florida lemons and from California lemons. Since the main ideas in this extension are similar to the ones presented in this paper, we omit the technical details and refer the interested reader to Babaioff (2001) for the details. Babaioff (2001) also discusses an extension of the model to the case that an agent is allowed to bid for multiple units in its market, by bidding the maximal quantity it is willing to trade and the price per unit.

A paper by Babaioff and Walsh (2004) extends our model to the case that each item is produced from a combination of several items of several different goods, but each good is still produced in exactly one market. They have presented a specific auction (unlike this work which presents generic protocols for supply chain formation) that is individually rational, incentive compatible, budget-balanced, yet highly efficient.

Similar results (IR, IC, BB and high efficiency) have been presented by Babaioff, Nisan, and Pavlov (2004) to a related model of Spatially Distributed Market. In this model a single good is traded in a set of independent markets, where shipment between markets is possible but incurs a publicly known cost (but in their model there are no strategic producers as in our model).

One interesting challenge is to stay with the unit-to-unit conversion model, but extend our model to allow a general directed a-cyclic supply chain graph topology. Yet another important challenge is to extend the model to a case where agents are not single-minded agents, and can submit exclusive or non exclusive bids in several markets.

## Acknowledgments

This research was supported by grants from the Israeli Ministry of Science, the Israeli Academy of Sciences and the USA-Israel Bi-national Science Foundation. The first author (Babaioff) was also supported by Yeshaya Horowitz Association.

We thank William Walsh, Daniel Lehmann, Hila Babaioff and the anonymous reviewers for their helpful comments.

## Appendix A. Problems with Chaining McAfee's Double Auction

**The Consistency Problem**: The following example of supply chain auction using McAfee's rule shows that not all double auction rules are consistent. Figure 14 shows the supply and demand curves in the three markets after the supply and demand curves propagation.

| **Lemon Market** | | **Squeezing Market** | | **Juice Market** | |
|---|---|---|---|---|---|
| **Supply** ($S^L$) | **Demand** ($D^L$) | **Supply** ($S^{L\rightarrow J}$) | **Demand** ($D^{L\rightarrow J}$) | **Supply** ($S^J$) | **Demand** ($D^J$) |
| **10** | 20 | **5** | 15 | 15 | **25** |
| **20** | 10 | **7** | -3 | 27 | **17** |

Figure 14: Chaining McAfee's DA example

The optimal trade quantity in this example is 1, as can be seen in Figure 14. McAfee's rule in the Lemonade Market set the trade size to 1, since $\frac{27+17}{2} = 22 \in [15, 25]$. On the other hand, if we use McAfee's rule on the Squeezing Market, trade reduction should be made since $\frac{7+(-3)}{2} = 2 \notin [5, 15]$ and the trade quantity is zero, in contradiction to the previous decision. We conclude that there exists a non-consistent DA rule that cannot be used by the Symmetric Protocol.

**The Revenue Problem**: The same example with the Pivot Protocol presented in Figure 15 shows that the fact that the DA rule has revenue surplus, does not ensure that the supply chain mechanism created by the Pivot Protocol using the DA rule has a revenue surplus as well.

| **Lemon Market** | | **Squeezing Market** | | **Juice Market** | |
|---|---|---|---|---|---|
| **Supply** ($S^L$) | **Demand** ($D^L$) | **Supply** ($S^{L\rightarrow J}$) | **Demand** ($D^{L\rightarrow J}$) | **Supply** ($S^J$) | **Demand** ($D^J$) |
| **10** | | **5** | | 15 | **25** |
| **20** | | **7** | | 27 | **17** |

$(V,q)=(17,1) \longleftarrow$ $\qquad$ $(V,q)=(22,1) \longleftarrow$

Figure 15: Pivot Protocol with McAfee's rule

In this example, McAfee's rule in the Lemonade Market sets the trade size to 1, since $\frac{27+17}{2} = 22 \in [15, 25]$. The buyer in this market should pay 22. By following the Pivot Protocol, the squeezer that has a bid of 5 should be paid $7 = min(7, 22 - 10)$ and the supplier in the supply market that has a bid of 10 should be paid $17 = min(17, 20)$. The total revenue is $22 - 7 - 17 = -2$ which means that there is a revenue deficit.

## Appendix B. Supply Chain Payments

In this appendix we prove Lemma B.1 which is used in the proof of Theorem 5.4.

**Lemma B.1.** *If the double auction rule used by the Pivot Protocol is IR, IC and non-discriminating, and $P_S$ is the "seller's price" in this double auction, then the total payment $P$ to a supplier and converters (one from each market) which create one unit of the final good satisfies $P_S \leq P \leq S^t_{q+1}$.*

*Proof.* By induction, each conversion market of $C^m$ to $C^{m+1}$ receives the pair $(P_S - \sum_{r=m+1}^{t-1} S_q^{r\to r+1}, q)$ from its demand market, and the payment to a winning converter is

$$min(P_S - \sum_{r=m+1}^{t-1} S_q^{r\to r+1} - S_q^m, S_{q+1}^{m\to m+1})$$

Similarly, the supply market $M^1$ receives the pair $(P_S - \sum_{r=1}^{t-1} S_q^{r\to r+1}, q)$, and the payment to a winning supplier is

$$min(P_S - \sum_{r=1}^{t-1} S_q^{r\to r+1}, S_{q+1}^1)$$

We conclude that the total payment to the converters and supplier of one unit is

$$P = \sum_{m=1}^{t-1} min(P_S - \sum_{r=m+1}^{t-1} S_q^{r\to r+1} - S_q^m, S_{q+1}^{m\to m+1}) + min(P_S - \sum_{r=1}^{t-1} S_q^{r\to r+1}, S_{q+1}^1)$$

First notice that $P \leq S^t_{q+1}$ since

$$P \leq \sum_{m=1}^{t-1} S_{q+1}^{m\to m+1} + S_{q+1}^1 = S_{q+1}^t$$

Secondly we should prove that $P_S \leq P$. The idea of the proof is that this holds if there was only one conversion from $C^1$ to $C^t$ (as we show in Lemma B.2), and by splitting a conversion to two consecutive conversions, the total payment to all the converters only grows (as we show in Lemma B.3).

**Lemma B.2.** *Assume that there was only one conversion market from $C^1$ to $C^t$, which means that for every $i$, $S_i^{1\to t} = \sum_{r=1}^{t-1} S_i^{r\to r+1}$ then*

$$P_S \leq P = min(P_S - S_q^1, S_{q+1}^{1\to t}) + min(P_S - S_q^{1\to t}, S_{q+1}^1)$$

*Proof.* Note that the double auction in the pivot market is the same as in the original auction, since the supply curve of the pivot market is the same, and therefore $P_S$ is the same.

We prove that this claim is true by checking the four possible cases:

1. if $P_S - S^1_q \leq S^{1\to t}_{q+1}$ and $P_S - S^{1\to t}_q \leq S^1_{q+1}$, then

$$P = (P_S - S^1_q) + (P_S - S^{1\to t}_q) = P_S + (P_S - (S^1_q + S^{1\to t}_q)) \geq P_S$$

since $P_S \geq S^1_q + S^{1\to t}_q = S^t_q$, because the DA rule is non-discriminating.

2. if $P_S - S^1_q \leq S^{1\to t}_{q+1}$ and $S^1_{q+1} \leq P_S - S^{1\to t}_q$, then

$$P = (P_S - S^1_q) + S^1_{q+1} = P_S + (S^1_{q+1} - S^1_q) \geq P_S$$

since $S^1_{q+1} \geq S^1_q$

3. if $S^{1\to t}_{q+1} \leq P_S - S^1_q$ and $P_S - S^{1\to t}_q \leq S^1_{q+1}$, then

$$P = S^{1\to t}_{q+1} + (P_S - S^{1\to t}_q) = P_S + (S^{1\to t}_{q+1} - S^{1\to t}_q) \geq P_S$$

since $S^{1\to t}_{q+1} \geq S^{1\to t}_q$.

4. if $S^{1\to t}_{q+1} \leq P_S - S^1_q$ and $S^1_{q+1} \leq P_S - S^{1\to t}_q$, then

$$P = S^{1\to t}_{q+1} + S^1_{q+1} \geq P_S$$

since $S^{1\to t}_{q+1} + S^1_{q+1} = S^t_{q+1} \geq P_S$, because the DA rule is non-discriminating.

So we have proven that $P \geq P_S$ in the case of one conversion market. □

**Lemma B.3.** *Assume that a conversion market from $C^k$ to $C^t$ is split into two conversion markets - one from $C^k$ to $C^{k+1}$ and the other from $C^{k+1}$ to $C^t$, then the payment for conversion of one unit from $C^k$ to $C^t$ can only grow, which means that*

$$P^{k\to t} \leq P^{k\to k+1} + P^{k+1\to t}$$

*where $P^{w\to z}$ is the cost of conversion of one $w$ unit to one $z$ unit. In other words, it is always true that*

$$min(P_S - S^k_q, S^{k\to t}_{q+1}) \leq min(P_S - S^{k+1\to t}_q - S^k_q, S^{k\to k+1}_{q+1}) + min(P_S - S^{k+1}_q, S^{k+1\to t}_{q+1})$$

*where $S^{k\to t}_i = \sum_{r=k}^{t-1} S^{r\to r+1}_i$*

*Proof.* Note that the double auction in the pivot market is the same as in the original auction, since the supply curve of the pivot market is the same, and therefore $P_S$ is the same.

We prove that this claim is true by checking the four possible cases:

1. if $P_S - S^{k+1\to t}_q - S^k_q \geq S^{k\to k+1}_{q+1}$ and $P_S - S^{k+1}_q \geq S^{k+1\to t}_{q+1}$, then

$$P^{k\to k+1} + P^{k+1\to t} = S^{k\to k+1}_{q+1} + S^{k+1\to t}_{q+1} = S^{k\to t}_{q+1} \geq min(P_S - S^k_q, S^{k\to t}_{q+1}) = P^{k\to t}$$

2. if $P_S - S_q^{k+1 \to t} - S_q^k \leq S_{q+1}^{k \to k+1}$ and $P_S - S_q^{k+1} \geq S_{q+1}^{k+1 \to t}$, then

$$P^{k \to k+1} + P^{k+1 \to t} = (P_S - S_q^{k+1 \to t} - S_q^k) + S_{q+1}^{k+1 \to t} =$$

$$(P_S - S_q^k) + (S_{q+1}^{k+1 \to t} - S_q^{k+1 \to t}) \geq P_S - S_q^k \geq P^{k \to t}$$

since $S_{q+1}^{k+1 \to t} \geq S_q^{k+1 \to t}$.

3. if $P_S - S_q^{k+1 \to t} - S_q^k \geq S_{q+1}^{k \to k+1}$ and $P_S - S_q^{k+1} \leq S_{q+1}^{k+1 \to t}$, then

$$P^{k \to k+1} + P^{k+1 \to t} = S_{q+1}^{k \to k+1} + (P_S - S_q^{k+1}) =$$

$$(P_S - S_q^k) + (S_{q+1}^{k \to k+1} - S_q^{k \to k+1}) \geq (P_S - S_q^k) \geq P^{k \to t}$$

since $S_{q+1}^{k \to k+1} \geq S_q^{k \to k+1}$.

4. if $P_S - S_q^{k+1 \to t} - S_q^k \leq S_{q+1}^{k \to k+1}$ and $P_S - S_q^{k+1} \leq S_{q+1}^{k+1 \to t}$, then

$$P^{k \to k+1} + P^{k+1 \to t} = (P_S - S_q^{k+1 \to t} - S_q^k) + (P_S - S_q^{k+1}) =$$

$$(P_S - S_q^k) + (P_S - S_q^{k+1} - S_q^{k+1 \to t}) \geq (P_S - S_q^k) \geq P^{k \to t}$$

since $P_S \geq S_q^{k+1} + S_q^{k+1 \to t} = S_q^t$.

So we have proven that the payment can only grow by splitting a conversion market. □

We conclude that the cost of one unit of $C^t$ is never smaller than $P_S$, since by the first lemma, it is true if there was only one conversion, and by induction and the second lemma, splitting the conversion market into all $t-1$ conversion markets can only increase the cost. □